\documentclass[%
 reprint,
 superscriptaddress,
 amsmath,amssymb,
 aps,
]{revtex4-2}

\usepackage[utf8]{inputenc}
\usepackage[T1]{fontenc}
\usepackage{graphicx}
\usepackage{epstopdf}
\usepackage{siunitx}
\usepackage{physics}
\usepackage[capitalize]{cleveref}
\usepackage[version=4]{mhchem}
\usepackage{lineno}
\usepackage{xcolor, soul}

\newcommand{\Cs}[1]{\textsuperscript{#1}{\text{Cs}}}

\begin{document}

\preprint{APS/123-QED}

\title{Optical Nanofiber Testbeds for Benchmarking Membrane-Waveguide Photonic Integrated Circuit Platforms toward On-Chip Quantum Inertial Sensing}

\author{Adrian Orozco}
\affiliation{Sandia National Laboratories, Albuquerque, New Mexico 87185, USA}
\affiliation{Department of Physics and Astronomy, University of New Mexico,  Albuquerque, NM 87106, USA}

\author{William Kindel}
\affiliation{Sandia National Laboratories, Albuquerque, New Mexico 87185, USA}

\author{Nicholas Karl}
\affiliation{Sandia National Laboratories, Albuquerque, New Mexico 87185, USA}

\author{Yuan-Yu Jau}
\affiliation{Sandia National Laboratories, Albuquerque, New Mexico 87185, USA}
\affiliation{Department of Physics and Astronomy, University of New Mexico,  Albuquerque, NM 87106, USA}

\author{Michael Gehl}
\affiliation{Sandia National Laboratories, Albuquerque, New Mexico 87185, USA}

\author{Grant Biedermann}
\affiliation{Department of Physics and Astronomy, University of Oklahoma, Norman, Oklahoma 73019, USA}

\author{Jongmin Lee}
\email{jlee7@sandia.gov}
\affiliation{Sandia National Laboratories, Albuquerque, New Mexico 87185, USA}
\affiliation{Department of Physics and Astronomy, University of New Mexico,  Albuquerque, NM 87106, USA}

\date{\today}

\begin{abstract}
Recent advances in cold atom interferometry with optical and magnetic atom guides have set the stage for quantum inertial sensors capable of operating in dynamic environments. In this work, we present three key innovations---evanescent-field (EF) atom guides, optical nanofiber testbeds, and membrane-waveguide photonic integrated circuit (PIC) platforms---to advance EF-guided atom interferometry. First, we demonstrate EF atom guides on optical nanofiber testbeds, which serve as performance benchmarks for our membrane-waveguide PIC platforms. Second, we achieve low-power (\SI{\sim 5}{mW}) guiding of freely moving, laser-cooled \Cs{133} atoms in two-color, traveling-wave EF optical dipole traps at the novel, heat-efficient magic wavelengths of 793 nm and 937 nm (i.e. "793/937-nm EF atom guides"). We designed and fabricated membrane-waveguide PIC platforms for these EF atom guides; in our prior work we showed that they safely handle up to 4--6$\times$ times the minimum trap power under vacuum and enable dense cold atom generation for efficient loading. Third, we verify preserved atomic coherence via microwave fields and EF-coupled Doppler-free Raman beams; to our knowledge, this is the first report of coherence fringes driven by co-propagating EF-coupled Raman beams with only \SI{150}{nW} of total optical power. By providing a direct comparison between optical nanofiber testbeds and membrane-waveguide PIC platforms, our results lay critical groundwork for the on-chip realization of EF-guided atom interferometry and the development of for fully integrated, low-SWaP (size, weight, and power) quantum accelerometers and gyroscopes.
\end{abstract}

\maketitle


\section*{Introduction}
\label{sec:intro}

Light-pulse atom interferometry \cite{Kasevich91} was developed for precision measurements of gravity \cite{Peters01} and for inertial sensing \cite{Gustavson00}. In recent years, atom interferometers have advanced significantly toward field applications \cite{McGuinness12, Cheiney18, Wu19, Lee22, Templier22, Stray22, Panda23, Seo26}. However, to facilitate the deployment of quantum inertial sensors in real-world scenarios,\cite{Bong19, Geiger20, Narducci22} atom interferometers must be miniaturized and ruggedized to accommodate dynamic motion. Specifically, ruggedization requires reliable transverse atomic confinement,\cite{Soh20} which researchers have demonstrated using both optical\cite{Close13, Katori17, Lan18, Zhan20} and magnetic\cite{Prentiss07, Fang17, Sackett20} atom guides. However, simultaneous ruggedization and miniaturization has not yet been demonstrated in a single device, or to the extent necessary for practical, real-world deployment.

\begin{figure}[b!]
\centering
\includegraphics[width=1\columnwidth]{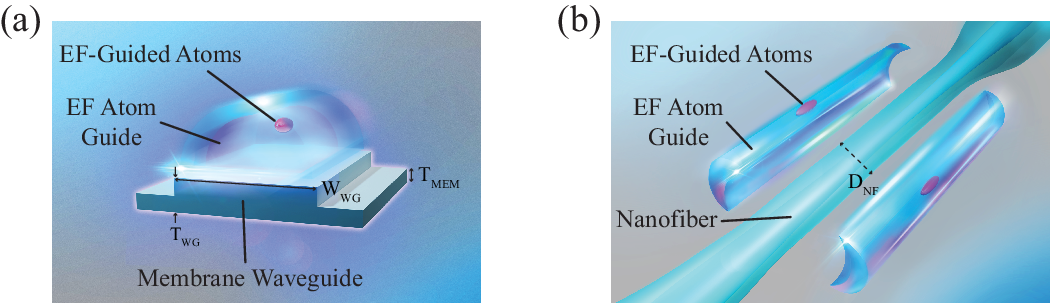}
\caption{Evanescent-field (EF) atom guides (i.e. EF optical dipole traps) realized using a membrane-waveguide photonic integrated circuit (PIC) platform and an optical nanofiber testbed. (a) Illustration of the EF atom guide, membrane waveguide, and EF-guided atoms; the waveguide geometry is characterized by the waveguide width ($\rm{W_{WG}}$), waveguide thickness ($\rm{T_{WG}}$), and membrane thickness ($\rm{T_{MEM}}$). (b) Illustration of the EF atom guide, optical nanofiber testbed, and EF-guided atoms; the nanofiber geometry is defined by the nanofiber diameter ($\rm{D_{NF}}$).}
\label{fig:concept}
\end{figure}

\begin{figure}[b!]
\includegraphics[width=1\columnwidth]{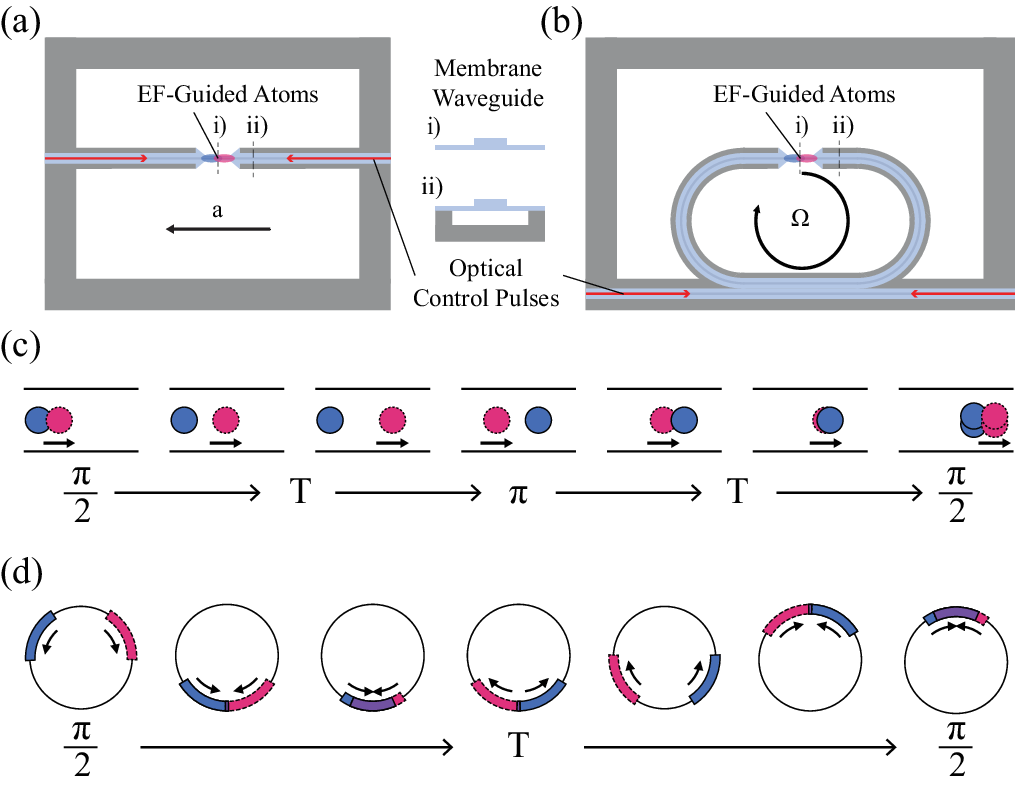}
\caption{On-chip quantum inertial sensors using EF atom guides implemented on membrane-waveguide PIC platforms, with EF-guided atom interferometry protocols employing EF-coupled Doppler-sensitive light-pulse sequences. (a) A quantum accelerometer using linear membrane waveguides. (b) A quantum gyroscope using racetrack membrane waveguides. (c) Linear acceleration measurement protocol involving three light pulses ($\frac{\pi}{2} \rightarrow$ T $\rightarrow \pi$ $\rightarrow$ T $\rightarrow \frac{\pi}{2}$), where T represents the interrogation time (equivalent to the free-evolution time in the Doppler-free Ramsey-echo sequence). (d) Angular velocity measurement protocol using two light pulses ($\frac{\pi}{2} \rightarrow$ T $\rightarrow \frac{\pi}{2}$). Both light-pulse sequences impart state-dependent photon recoils to the EF-guided atoms, enabling precise measurements of linear acceleration and angular velocity. The two atomic states are denoted by red and blue solid circles and annular sectors.}
\label{fig:protocol}
\end{figure}

Sub-micron optical waveguides offer a promising route to bridge that gap. Both evanescent-field (EF) optical traps \cite{Balykin04} and magnetic traps \cite{Schneeweiss14, Vylegzhanin25} have been proposed to localize atoms within a few hundred nanometers of the nanofiber surface. In this work, we focus on integrating optical-waveguide-based evanescent-field (EF) atom guides (i.e. EF optical dipole traps, see \cref{fig:concept}) on compact, robust photonic integrated circuits (PICs) to achieve EF-guided atom interferometry (\cref{fig:protocol}) with substantially reduced size, weight, and power (SWaP). A crucial advantage of this approach is that EF atom guides confine light in a much smaller mode area than traditional free-space optical dipole traps, significantly reducing the required optical power.\cite{Ovchinnikov91, Miller93, Grimm20, Boshier22} However, effective integration must overcome a substantial technical hurdle: generating a large number of atoms around the EF atom guide while ensuring adequate thermal management in a vacuum environment.

\begin{table*}
\centering
\begin{center}
\begin{tabular}{cccccc}
  \hline
  Waveguide Type & Wavelengths and Optical Powers & Light Pol. & Total Power & Trap Depth& Trap Dir. (Surface Dist.) \\ 
  \hline
  Membrane Waveguides & ($P_{793 nm}$, $P_{937 nm}$) = (3.27, 2.73) \SI{}{mW} & Lin$\parallel$Lin & \SI{6}{mW} & \SI{350}{\micro \kelvin} & Light Pol. $\perp$ (\SI{120}{nm}) \\ 
  \hline
  Optical Nanofibers & ($P_{793 nm}$, $P_{937 nm}$) = (6.8, 3.9) \SI{}{mW} & Lin$\parallel$Lin & \SI{10.7}{mW} & \SI{350}{\micro \kelvin} & Light Pol. $\parallel$ (\SI{260}{nm}) \\ 
  \hline
  Optical Nanofibers & ($P_{685 nm}$, $P_{937 nm}$) = (25, 2.5) \SI{}{mW} & Lin$\parallel$Lin & \SI{27.5}{mW} & \SI{350}{\micro \kelvin} & Light Pol. $\parallel$ (\SI{260}{nm}) \\ 
  \hline
\end{tabular}
\caption{Comparison of evanescent-field (EF) atom guides (i.e. EF optical dipole traps) for \Cs{133} atoms on: (Case 1) a membrane-waveguide photonic integrated circuit (PIC) platform at 793/937 nm, (Case 2)  an optical nanofiber testbed at 793/937 nm, and (Case 3) an optical nanofiber testbed at 685/937 nm. Abbreviations: Light Pol., polarization direction of the blue- and red-detuned trapping beams; Trap Dir., orientation of atom trapping relative to the light polarization; Surface Dist., atom-surface distance.}
\label{tab:trap_comp}
\end{center}
\end{table*}

Traditional optical waveguides on opaque substrates\cite{Rolston15, Harding25, Zhou25} are simple to fabricate, offer efficient heat dissipation, and support higher optical power in vacuum. In this case, cold atom generation near the waveguides is hindered by atom-surface collisions. In another case, waveguides suspended on transparent membranes\cite{Ayi-Yovo20, Lee21, Gehl21, Ovchinnikov25} (i.e. membrane waveguides) facilitate dense cold atom generation around the structure but suffer from limited heat dissipation, leading to heat accumulation and elevated temperatures. In both cases, achieving efficient EF atom guides requires a radically different approach. To meet this need (enabling dense cold atom generation, effective heat dissipation, and low optical power operation), we developed a novel membrane-waveguide PIC platform (\cref{fig:concept}a). It comprises a membrane optical waveguide suspended over an aperture in a transparent membrane anchored to a silicon substrate (see \cref{fig:WG_Concept}a $\&$ \cref{fig:fab_device}).\cite{Lee21, Gehl21}

Experimental tests demonstrate that the membrane waveguide can support optical powers of 20--\SI{30}{mW} before fracturing, easily fulfilling the requirements for EF atom guiding (see Case 1 of \cref{tab:trap_comp} for the 793/937 nm configuration). Using a membrane magneto-optical trap (MOT) that leverages a larger capture volume formed by six laser-cooling beams transmitted through the transparent membrane,\cite{Lee21} we generated between $10^4$--$10^5$ sub-Doppler-cooled atoms (\SI{\sim 10}{\micro \kelvin}) at the small open hole of the membrane-waveguide PIC platforms. 

Despite these advancements, direct atom trapping relying solely on EF-coupled beams on a PIC platform remains elusive, prompting researchers to employ additional external free-space trapping beams. \cite{Hung24} Therefore, reliable optical nanofiber testbeds (\cref{fig:concept}b) that are capable of handling high optical powers (> \SI{100}{mW}) under vacuum are essential for exploring alternative designs and light configurations. Most nanofiber testbeds have been demonstrated for EF optical lattices that employ standing evanescent waves.\cite{Vestch10, Goban12, Reitz13, Lee15, Polzik17, Meng18, Aoki19, Corzo19, Chromaic22, Kestler23} In this work, we use optical nanofiber testbeds that employ traveling evanescent waves to form EF atom guides, which confine atoms transversely in two dimensions while allowing free motion along the fiber.

\cref{tab:trap_comp} presents a performance comparison of EF atom guides for \Cs{133} atoms ($\lambda$ = 852$\text{-}$nm) on a membrane-waveguide PIC platform at 793/937 nm (Case 1) and two optical nanofiber testbeds at 793/937 nm (Case 2) and 685/937 nm (Case 3). Compared to the standard magic wavelengths (685/973 nm), we kept the red-detuned light at 937 nm, but chose a blue-detuned light at the magic-wavelength of 793 nm, which lies closer to resonance (\cref{tab:trap_comp} $\&$  \cref{fig:LS_Pol}). Using this latter configuration, we were able to demonstrate EF atom guiding on an optical nanofiber testbed with a total optical power of \SI{\sim 5}{mW} (\cref{fig:WG_Concept}b). These low-power 793/937-nm wavelengths would also enable simpler thermal management in vacuum environments for membrane-waveguide PIC platforms.\cite{Gehl21}

To validate the performance of the 793/937-nm EF atom guides on an optical nanofiber testbed and to benchmark them on membrane-waveguide PIC platforms, we assessed the atomic coherence of the EF-guided atoms with both microwave fields and EF-coupled Raman beams. In the Ramsey phase-scan measurement, the EF-coupled Raman beams (excited in the HE$_{11}$ cladding-air mode) drive Doppler-free Raman transitions via the the light-pulse sequence $\frac{\pi}{2} \rightarrow$ T $\rightarrow$ $\pi$ $\rightarrow$ T $\rightarrow \frac{\pi}{2}$ ($\delta \phi$) with a relative phase shift $\delta \phi$, producing sinusoidal fringe contrasts as $\delta \phi$ is scanned. Subsequent work will test a protocol using Doppler-sensitive Raman transitions that impart state-dependent photon recoils to EF-guided atoms, generating matterwave interference for linear acceleration measurements along the EF guides (\cref{fig:protocol}a$\&$c).

To benchmark the 793/937-nm EF atom guides that were originally designed for membrane-waveguide PIC platforms, we used optical nanofiber testbeds to measure the atom number, lifetime, and atomic coherence of EF-guided atoms under both microwave fields and EF-coupled Doppler-free Raman beams. To our knowledge, this is the first demonstration of coherence fringes in EF atom guides driven by co-propagating EF-coupled Raman beams using only \SI{150}{nW} of optical power. Operating with just \SI{5}{mW} of total power, the nanofiber testbeds support stable EF atom guiding and yield clear Rabi oscillations and Ramsey fringes, thereby validating  the low-power performance of the 793/937-nm EF atom guides on membrane-waveguide PIC platforms. Finally, we propose a Doppler-sensitive Raman protocol for EF-guided atom interferometry, paving the way to compact, low-SWaP quantum inertial sensors.

\section*{Results}

Recent advancements in PIC platforms \cite{Stern13, Rolston13a, Lukin14, Kimble15, Rolston15, Stievater16, Pfau18, Hung19, Hackermueller20, Ayi-Yovo20, Kimble20, Lee21, Gehl21, Hung24} have improved scalability and design flexibility for atom-light interactions and atom-trapping configurations. More critically, our membrane-waveguide PIC platforms enable dense cold atom generation for efficient loading into EF atom guides (\cref{fig:concept}), effective heat dissipation, and reduced optical power requirements, while also supporting alternative geometries that are optimized for different inertial measurements, including both linear (\cref{fig:protocol}a$\&$c) and racetrack (\cref{fig:protocol}b$\&$d) designs. We designed and fabricated these membrane-waveguide PIC platforms (\cref{fig:fab_device}) to accommodate 793/937-nm EF atom guides (\cref{fig:WG_Concept}a $\&$ \cref{fig:LS_Pol}) and then tested the EF atom guides on optical nanofiber testbeds (\cref{fig:WG_Concept}b $\&$ \cref{fig:setup}). We characterized the atom number and lifetime of EF-guided atoms (\cref{fig:lifetime_793_937}) and validated the atomic coherence with both microwave fields (\cref{fig:coherence_793_937_Microwave}) and EF-coupled Doppler-free Raman beams (\cref{fig:coherence_793_937_DF_Raman}). We also characterized microwave coherence for a 685/937-nm EF atom guide (\cref{fig:coherence_685_937_Microwave}). The phase-scan measurement of Ramsey interferometry with an echo, utilizing EF-coupled Doppler-free Raman beams, closely resembles the light-pulse sequence used in EF-guided atom interferometry with EF-coupled Doppler-sensitive Raman beams. These advancements are expected to accelerate the realization of real-world, deployable on-chip quantum inertial sensors,\cite{Lee24a, Lee24b, Lee25} as illustrated in \cref{fig:protocol}.

\subsection*{Design and Optimization of Membrane-Waveguide Photonic Integrated Circuit (PIC) Platforms for Dense Cold Atom Generation, Efficient Loading, and Thermal Management}

\begin{figure}[t!]
\includegraphics[width=1\columnwidth]{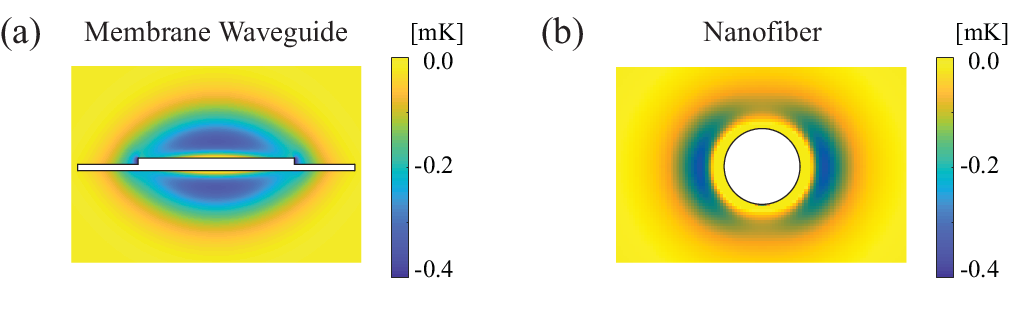}
\caption{Optical potential of the EF atom guide on two platforms: (a) a membrane-waveguide PIC platform and (b) an optical nanofiber testbed. In both cases, the EF atom guide is formed by blue-detuned (793 nm) and red-detuned (937 nm) traveling evanescent waves, yielding a maximum trap depth of approximately \SI{350}{\micro \kelvin} at the 852 nm D2 transition of \Cs{133} atoms (the first and second rows of \cref{tab:trap_comp}).}
\label{fig:WG_Concept}
\end{figure}

Membrane-waveguide PIC platforms enhance the design and scalability of EF atom guides, which are crucial for EF-guided atom interferometry (\cref{fig:protocol}). These devices utilize two-color traveling evanescent waves to optimize repulsive (blue-detuned) and attractive (red-detuned) potentials within the van der Waals potential. By integrating suspended membrane waveguide structures with either a membrane MOT (in the infinity design) or a conventional MOT (in the hybrid-needle design),\cite{Lee21} dense cold atom generation for efficient loading into EF atom guides is facilitated. Optimizing the key design parameters (waveguide width $\rm{W_{WG}}$, waveguide thickness $\rm{T_{WG}}$, and membrane thickness $\rm{T_{MEM}}$) maximizes the trap potential per optical power when $\rm{T_{MEM}}$ is thinner than $\rm{T_{WG}}$. For effective cooling, optimal $\rm{T_{MEM}}$ ensures maximum transmission of circularly polarized beams. Our simulation based on the membrane-waveguide PIC platform predicts that 793/937-nm EF atom guides, operated in a lin$\parallel$lin polarization configuration, require a total optical power of \SI{6}{mW} to achieve a maximum trap depth of \SI{350}{\micro \kelvin} (Case 1 of \cref{tab:trap_comp} $\&$ \cref{fig:WG_Concept}a) with ($P_{793}$, $P_{937}$) = (\SI{3.27}{mW}, \SI{2.73}{mW}). The EF-guided atoms are located within \SI{120}{nm} of the waveguide surface, perpendicular to the polarization direction. The device parameters are: $\rm{W_{WG}} =$ \SI{1.6}{\micro m}; $\rm{T_{WG}} =$ \SI{100}{nm}, and $\rm{T_{MEM}} =$ \SI{50}{nm}. Note that, for the fundamental mode, the trap depth in the lateral (polarization) direction is shallower than that in the vertical direction. To compensate for this reduced lateral confinement, the waveguide’s refractive index or thickness can be varied locally. Alternatively, two-mode ($\rm TE_{00}$/$\rm TE_{01}$) and dual‐resonant schemes, \cite{Ovchinnikov25} which employ higher-order modes and larger wavelength separations, can enhance transverse confinement but may incur inter‐mode conversion and higher bending losses.

\subsection*{Developments of Membrane-Waveguide PIC Platforms for EF-Guided Atom Interferometry}

\begin{figure}[t!]
\includegraphics[width=1\columnwidth]{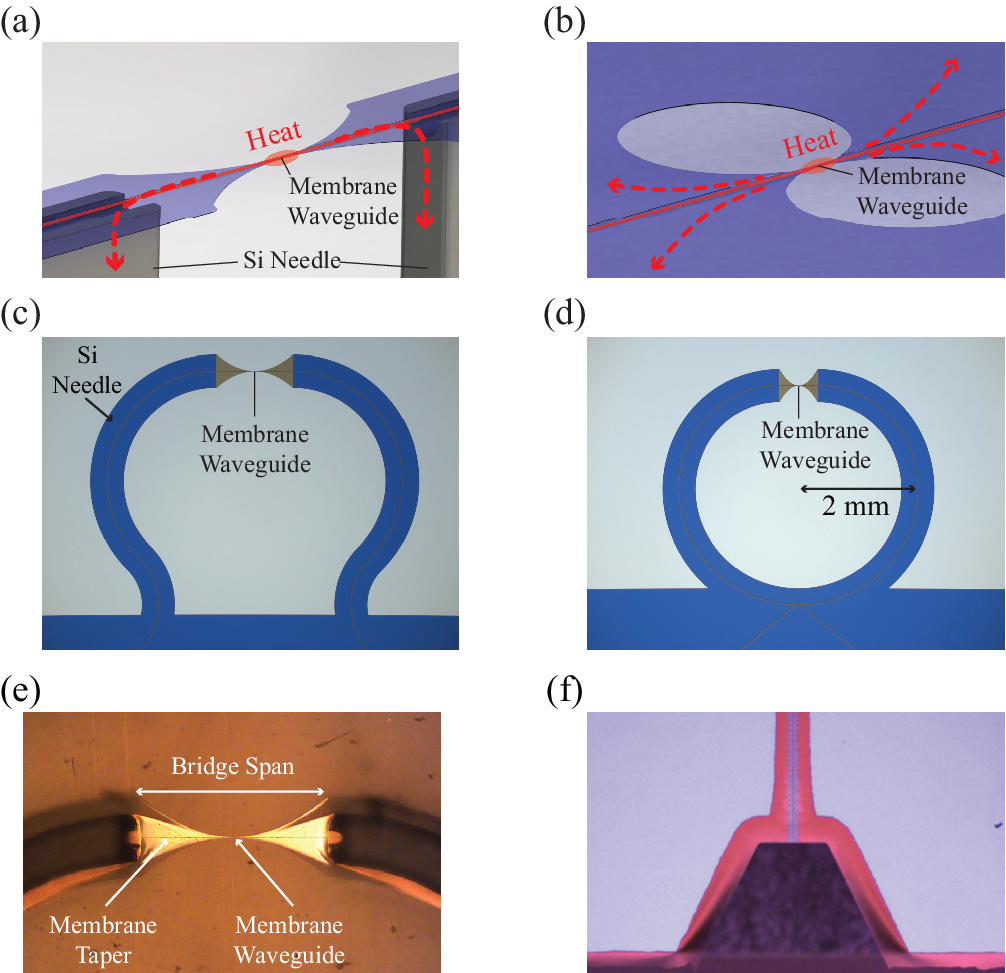}
\caption{Alumina ($\rm{Al_2O_3}$)  membrane-waveguide PIC platforms in hybrid-needle and infinity designs, and specialized geometries for area-enclosed matterwave interferometers. (a) Hybrid-needle design: a suspended membrane waveguide spans two silicon (Si) needles with undercut trenches. (b) Infinity design: a membrane waveguide is defined between two adjacent holes whose edges nearly touch. (c) Omega-shaped platform: semi-enclosed loop for circular atom guiding (\SI{1.4}{mm} span). (d) Ring-shaped platform: fully enclosed Sagnac loop (\SI{700}{\micro m} span). Panels (c) and (d) show the devices before $\rm{XeF}_{2}$ release: off-white regions become open gaps, blue regions remain as alumina membrane and membrane waveguides above the Si structures, and brown regions form suspended membrane. (e) Released omega platform mounted on a glass slide, featuring a membrane waveguide taper for circular atom guiding and improved heat dissipation. The inner rectangular opening area (\SI{9}{mm}$\times$\SI{9.6}{mm}) enables dense cold atom generation near the EF atom guide. (f) Fiber trench cutout with rounded corners: protects the waveguide facet during $\rm{XeF}_{2}$ etch, maintain facet tension, and support efficient free-space coupling.}
\label{fig:fab_device}
\end{figure}

Our alumina ($\rm{Al_2O_3}$) membrane-waveguide PIC platforms integrate linear, circular, and arbitrary-shaped EF atom guides for collective atom-light interactions in a manufacturable form. By leveraging sub-micrometer EF modes, these platforms offer compactness, robustness, and energy efficiency for on-chip EF-guided atom interferometry. As shown in \cref{fig:fab_device}, the Sandia platforms provide: (1) dense sub-Doppler-cooled atom generation in the vicinity of the membrane waveguide, (2) efficient loading of those atoms into EF atom guides, and (3) effective heat dissipation through the transparent alumina membrane and silicon substrate.

In the hybrid-needle design (\cref{fig:fab_device}a),  the membrane waveguide is suspended between two undercut Si-needle pillars. In the infinity design (\cref{fig:fab_device}b), two closely spaced membrane apertures form a narrow bridge for the membrane waveguide. Both geometries sustain up to 20--\SI{30}{mW} \cite{Gehl21} of optical power in vacuum, sufficient for EF atom guiding for \Cs{133} atoms. We have also demonstrated omega-shaped (\cref{fig:fab_device}c) and ring-shaped (\cref{fig:fab_device}d) platforms tailored for angular velocity sensing.

EF-guided atom interferometry on these PICs delivers several advantages over free-space optical dipole traps: reduced system size, intrinsic optical alignment, and a three-orders-of-magnitude reduction in required optical power. This power saving enables scalable, multi-axis inertial measurements with large radial confinement (\SI{> 1e4}{m/s^2}) and trap frequencies of \SI{\approx 100}{kHz}. Additionally, state-dependent atom detection is inherently EF-coupled.

For a gyroscope configuration, we designed a ring resonator (\cref{fig:fab_device}d) with a minimum radius of \SI{2}{mm}, total length \SI{13.2}{mm} and enclosed area \SI{14.2}{mm^2}. Simulations indicate that a 200-nm coupling gap over \SI{13}{\micro m} yields optimal phase matching; our current prototype uses a 350-nm coupling gap and an $\sim$50-\SI{}{\micro m} length, resulting in sub-optimal phase matching.

To improve fiber coupling, we introduced a rounded-corner membrane cutout, which shields the waveguide facet during $\rm{XeF}_{2}$ release (\cref{fig:fab_device}f) and maintains tension for high-efficiency coupling. We also estimated the coupling efficiencies of inverse tapering at the input facet to be \SI{33}{\%} at 793 nm, \SI{36}{\%} at 852 nm, and \SI{42}{\%} at 937 nm, noting that additional improvements may be possible by further narrowing the taper tip.

\subsection*{793/937-nm EF Atom Guides as Magic Wavelengths for \Cs{133} Atoms}

EF atom guides on optical nanofiber testbeds leverage traveling evanescent waves to generate both repulsive (blue-detuned) and attractive (red-detuned) potentials. In combination with van der Waals interactions, these light-induced potentials produces a stable potential minimum near the fiber surface. The total AC Stark (light) shift is $\Delta E_{\rm AC} = \sum_i \hbar\Omega^2_i/4\Delta_i$, where $i$ indexes atomic transitions, $\hbar$ is the reduced Planck's constant, $\Delta_i$ is the detuning, and $\Omega_i$ is the Rabi frequency. Unlike free-space atom guides,\cite{Miller93, Grimm20} which focus a single-color beam on a small spot, EF atom guides enable one-dimensional atom confinement with a long, uniform potential and greatly enhanced atom-light interactions. Because EF atom guides have much smaller EF-coupled mode area, they enable low-power atom trapping and low-powered Raman beams, reducing the required optical power by three to five orders of magnitude.

\begin{figure}[b!]
\includegraphics[width=1\columnwidth]{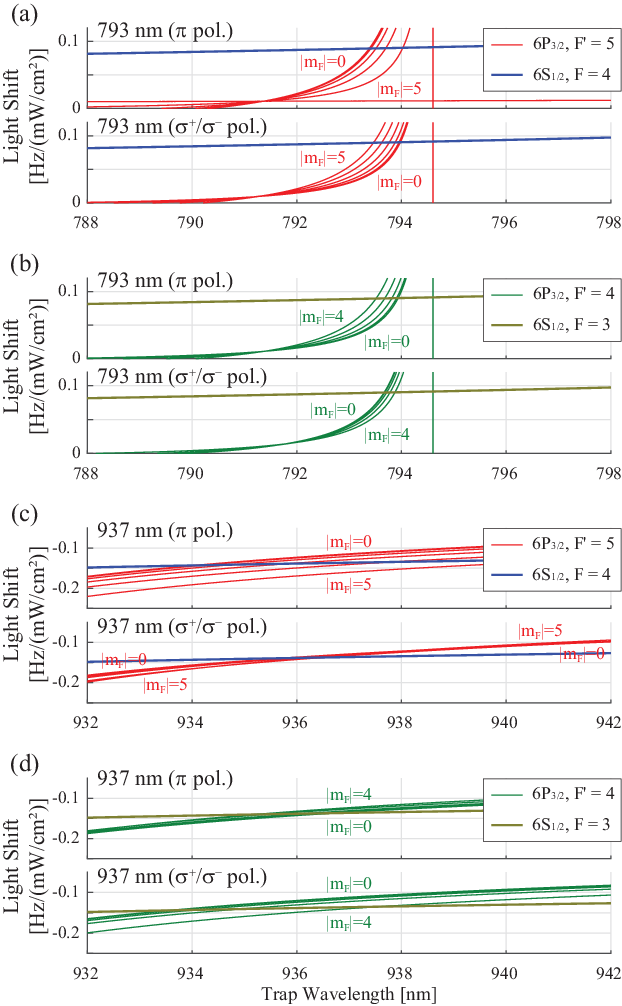}
\caption{Calculation of the light shift (LS) for \Cs{133} energy levels as a function of trap wavelength (nm) in vacuum for $\pi$ and $\sigma^{+}/\sigma^{-}$ polarizations: (a) LS for the 6S$_{1/2}$ $\ket{F=4}$ and 6P$_{3/2}$ $\ket{F'=5}$ transitions with 793-nm light. (b) LS for the 6S$_{1/2}$ $\ket{F=3}$ and 6P$_{3/2}$ $\ket{F'=4}$ transitions with 793-nm light. (c) LS for the 6S$_{1/2}$ $\ket{F=4}$ and 6P$_{3/2}$ $\ket{F'=5}$ transitions with 937 nm light. (d) LS for the 6S$_{1/2}$ $\ket{F=3}$ and 6P$_{3/2}$ $\ket{F'=4}$ transitions with 937 nm light.}
\label{fig:LS_Pol}
\end{figure}

For the 6S$_{1/2}$-to-6P$_{3/2}$ D2 transition of \Cs{133} atoms, the 793/937-nm wavelengths are magic wavelengths, \cite{Kimble12} minimizing the light-shift impact of the transition and enabling efficient loading of laser-cooled atoms into EF atom guides without changing the laser-cooling detuning. To confirm their effectiveness, we calculated the light shifts as a function of wavelength (\cref{fig:LS_Pol}), focusing on the 6S$_{1/2}$ and 6P$_{3/2}$ transitions for $\pi$ and $\sigma^{+}/\sigma^{-}$ polarizations. 

Using two-color, traveling evanescent waves with lin$\parallel$lin polarization in an optical nanofiber, we find that achieving a trap depth of \SI{350}{\micro \kelvin} at the potential minimum (\SI{\sim 260}{nm} from the nanofiber surface) requires $P_{793}$ = \SI{6.8}{mW} and $P_{937}$ = \SI{3.9}{mW} (see Case 2 of \cref{tab:trap_comp}). We implemented these 793/937-nm EF atom guides with \SI{\sim 5}{mW} and verified atomic coherence using microwave fields and EF-coupled Doppler-free Raman beams. The coherence times of the EF-guided atoms---without axial confinement---match those observed in EF optical lattices.\cite{Vestch10, Goban12, Reitz13, Lee15, Polzik17, Meng18, Aoki19, Corzo19, Chromaic22, Kestler23} This proof-of-concept reduces the required optical power dramatically, demonstrating its suitability for membrane-waveguide PIC platforms, which offer improved heat dissipation in vacuum.

\subsection*{Experimental Setup and Process for EF Atom Guides on an Optical Nanofiber}

\begin{figure}[t!]
\centering
\includegraphics[width=1\columnwidth]{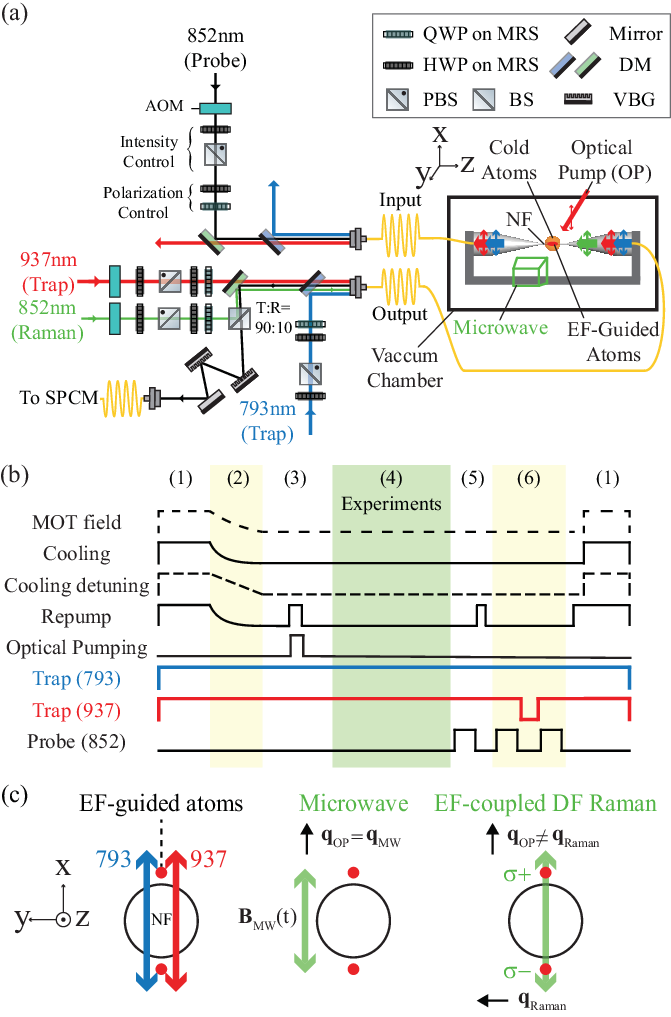}
\caption{Experimental setup and procedures. (a) Experimental setup for EF-guided \Cs{133} atoms (852 nm) on an optical nanofiber, utilizing blue- and red-detuned traveling evanescent waves. Key components include a single photon counting module (SPCM), a quarter-wave plate (QWP), a half-wave plate (HWP), a motorized rotation stage (MRS), a polarizing beam splitter (PBS), a dichroic mirror (DM), and a volume Bragg grating (VBG). (b) Experimental steps. (c) Experimental configurations for EF-guided atoms, microwave coherence measurements, and EF-coupled Raman coherence measurements. The quantization axes are denoted as $\mathbf{q}_{\rm OP}$ for optical pumping, $\mathbf{q}_{\rm MW}$ for microwave fields, and $\mathbf{q}_{\rm Raman}$ for Raman beams.}
\label{fig:setup}
\end{figure}

The experimental setup is shown in \cref{fig:setup}. The EF atom guide utilizes blue- and red-detuned beams combined and separated by dichroic mirrors, with polarization and intensity optimized using motorized rotation stages. Volume Bragg gratings filter the 852 nm probe beam, which is detected by a single photon counting module (SPCM).

The six experimental steps are shown in \cref{fig:setup}b. In step (1), atoms are trapped and cooled in a six-beam MOT, forming an atomic cloud that overlaps the waist of the optical nanofiber. The cooling beam is red-detuned by \SI{10}{MHz} from the $\ket{F = 4} \rightarrow \ket{F' = 5}$ transition with $I_{\rm cool}$ = \SI{12.3}{mW/cm^2}, and the repump beam is resonant on the $\ket{F = 3} \rightarrow \ket{F' = 4}$ transition with $I_{\rm repump}$ = \SI{1.1}{mW/cm^2}. In step (2), sub-Doppler cooling reduces the temperature to \SI{7}{\micro \kelvin} by lowering the cooling beam intensity by a factor of 1000 and increasing the detuning to \SI{60}{MHz}, \cite{Steane91} while minimizing collisional blockade.\cite{Rolston13a} In step (3), we optically pump the atoms into the dark $\ket{F = 4, m_F = 0}$ state using a $\pi$-polarized beam resonant with the $\ket{F = 4, m_F=0} \rightarrow \ket{F' = 4, m_F = 0}$ transition, preparing them for subsequent coherence measurements (\cref{fig:coherence_793_937_Microwave}, \cref{fig:coherence_793_937_DF_Raman}, \cref{fig:coherence_685_937_Microwave}). This step is omitted for atom number and lifetime measurements (\cref{fig:lifetime_793_937}), which consider all Zeeman sublevels of the $\ket{F = 4}$ manifold. In step (4), we perform the physics experiments (green region), including Rabi and Ramsey coherence protocols on the $\ket{F = 4, m_F = 0} \leftrightarrow \ket{F = 3, m_F = 0}$ transition  (\cref{fig:coherence_793_937_Microwave}, \cref{fig:coherence_793_937_DF_Raman}, \cref{fig:coherence_685_937_Microwave}).  In steps (5)--(6), we detect atomic populations for coherence measurements using a three-pulse, 852-nm absorption probe sequence following the state preparation in step (3): step (5) provides an initial, state-selective probe, followed by a repump pulse that transfers any $\ket{F=3}$ population into $\ket{F=4}$. In step (6), the first probe measures the total $\ket{F=4}$ population with the 937-nm trapping beam on, and the second probe measures the background absorption with the trapping beam off. For atom number and lifetime measurements, we omit step (3) and skip the step (5), using only the two probe pulses in step (6) to measure the total $F = 4$ population both with and without the 937-nm trapping beam.

\subsection*{Atom Number and Lifetime Measurements in 793/937-nm EF Atom Guides on an Optical Nanofiber}

\begin{figure}[t!]
\includegraphics[width=0.9\columnwidth]{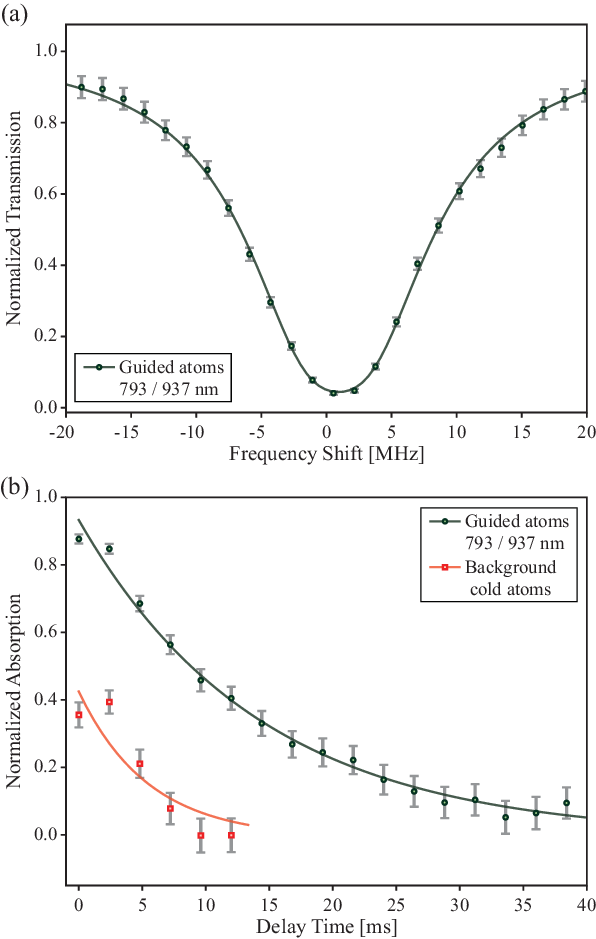}
\caption{Atom number and lifetime measurements of EF-guided atoms in 793/937-nm EF atom guides. (a) In a 793/937-nm EF atom guide with a total power of \SI{5.27}{mW} ($P_{793}$ = \SI{3.84}{mW}, $P_{937}$ = \SI{1.43}{mW}), the average number of EF-guided atoms (N$_{\rm Trap}$) is calculated to be 45\SI{\pm 2}{} (green circle) from the probe transmission. The frequency shift due to the EF atom guide is \SI{1.4}{MHz}. (b) The lifetime of EF-guided atoms ($\tau_\text{Trap}$) is estimated to be 14.3\SI{\pm 1.0}{ms} from the probe absorption, compared to the lifetime of background cold atoms at 5.5\SI{\pm 1.0}{ms}. Each data is averaged over 10 data points. All measurements were taken with an EF-coupled probe (\Cs{133}, 852 nm).}
\label{fig:lifetime_793_937}
\end{figure}

Using absorption spectroscopy, we measured both EF-guided atoms and background cold atoms (formed around an optical nanofiber with the guiding beams off) after sub-Doppler cooling with an EF-coupled probe (\Cs{133}, 852 nm) resonant on the $\ket{F = 4} \rightarrow \ket{F' = 5}$ transition, which measures the absorption of atoms distributed in the $\ket{F = 4}$ state (\cref{fig:lifetime_793_937}). Generally, a larger number of atoms in the guide leads to increased resonant absorption.
 
The total optical depth is defined as $\rm{OD} = -\ln(T)$, where $\rm{T}$ is the transmission of the absorption probe. In this experiment (\cref{fig:lifetime_793_937}), transmission data were fitted to a Lorentzian function to calculate $\rm{OD}$, expressed as $\rm{T}(\omega) = \exp[- \frac{\rm{OD}}{1 + 4 (\omega - \omega_0)^2/\Gamma^2}]$. Here, $\omega_0$ is the atomic resonant frequency, and $\Gamma$ is the linewidth of atomic transition ($= 2\pi \cdot$ \SI{5.2}{MHz}). The optical depth of EF-guided atoms is denoted as $\rm{OD}_{\rm Trap}$. The number of atoms is calculated as $\rm{N}_{\rm Trap} = \rm{OD}_{\rm Trap}/\rm{OD}_1$ (\cref{fig:lifetime_793_937}), where $\rm{OD}_1$ (\SI{\sim 0.08}{}) is the single-atom optical depth. Given a total power of \SI{5.27}{mW} ($P_{793}$ = \SI{3.84}{mW}, $P_{937}$ = \SI{1.43}{mW}; see Case 2 of \cref{tab:trap_comp}), we obtained N$_{\text{Trap}}$ = 45\SI{\pm 2}{} EF-guided atoms on an optical nanofiber testbed (green circle, \cref{fig:lifetime_793_937}a).

To investigate the lifetime of EF-guided atoms on the optical nanofiber, we varied the delay between loading the atoms into the EF guide and taking the transmission measurement, fitting the data points ($\rm{OD}$) to an exponentially-decaying function. For the 793/937-nm EF atom guide, the 1/e lifetime of EF-guided atoms is $\tau_\text{Trap}$ = 14.3\SI{\pm 1.0}{ms} (green circle, \cref{fig:lifetime_793_937}b), compared to $\tau_{\rm Bg}$ = 5.5\SI{\pm 1.0}{ms} for background cold atoms (red square). Furthermore, the lifetime of the EF-guided atoms is significantly longer than that of the background cold atoms (approximately \SI{5}{ms} without EF guiding). 

\subsection*{Atomic Coherence Measurements Using Microwave Fields in 793/937-nm EF Atom Guides on an Optical Nanofiber}

\begin{figure}[t!]
\includegraphics[width=1\columnwidth]{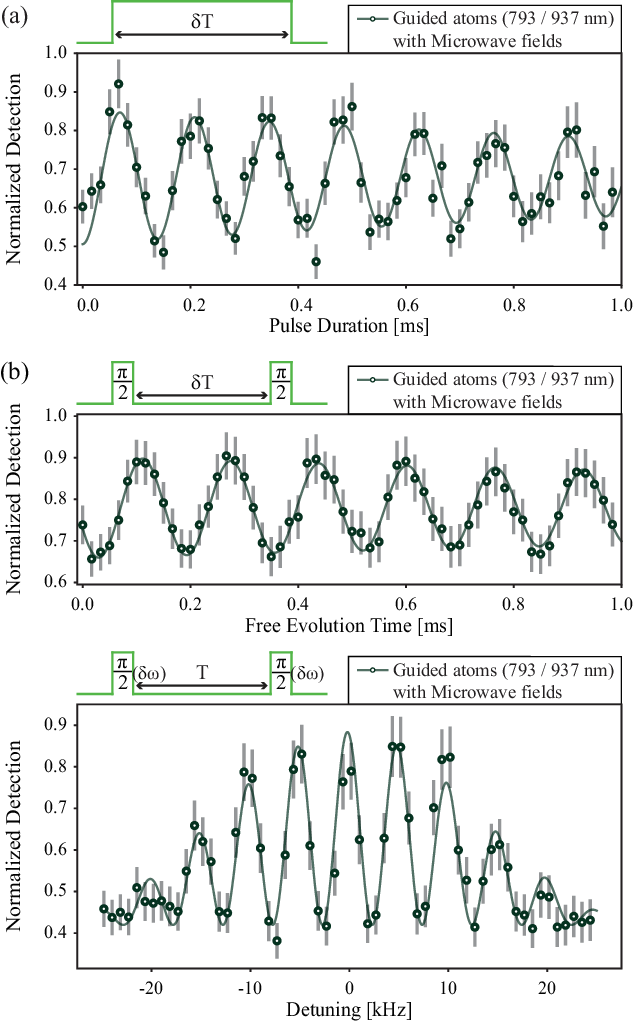}
\caption{Measurements of atomic coherence in EF-guided atoms using microwave fields. The optical nanofiber testbed for EF atom guides employs light wavelengths of 793 nm and 937 nm. (a) Rabi oscillation measurement; the $1/e$ decay time is $\tau_{1/e}$ = 1.8\SI{\pm 0.7}{ms}. (b) Ramsey coherence measurements. (i) Time-scan Ramsey measurement, where the Ramsey sequence is $\frac{\pi}{2} \rightarrow$ $\delta$T $\rightarrow \frac{\pi}{2}$, the coherence time is $\tau_2^*$ = 3.2\SI{\pm 1.1}{ms}, and the $\pi$ pulse time of the microwave is \SI{80}{\micro s}. (ii) Frequency-scan Ramsey measurement, where the Ramsey sequence is $\frac{\pi}{2}$ ($\delta\omega$) $\rightarrow$ T $\rightarrow$ $\frac{\pi}{2}$ ($\delta\omega$) and the free-evolution time is T = \SI{150}{\micro s}. In all panels each data point is the average of 50 measurements.}
\label{fig:coherence_793_937_Microwave}
\end{figure}


We first investigated the atomic coherence of EF-guided atoms by driving the atomic clock transition, $\ket{F = 3, m_F = 0} \rightarrow \ket{F = 4, m_F = 0}$, using a microwave horn. As shown in \cref{fig:setup}b, the physics measurement (green region) follows the initial state preparation. The microwave experiment includes optical pumping and establishing a quantization axis of \SI{3}{G}. To drive microwave Rabi oscillations (\cref{fig:coherence_793_937_Microwave}a), the microwave field is nearly resonant with the microwave atomic clock transition (\SI{\sim 9.192}{GHz}), including the light shift; increasing the microwave pulse length reveals the coherent Rabi oscillation. 
For the time-scan Ramsey measurement [\cref{fig:coherence_793_937_Microwave}b(i)], we extend the free-evolution time to $T = \rm{T}_0 + \delta T$ at the resonant microwave frequency, enabling the measurement of atom-interferometric fringes via the pulse sequence $\frac{\pi}{2}$ $\rightarrow$ $\rm{T}_0 + \delta T$ $\rightarrow$ $\frac{\pi}{2}$ and yielding a coherence time of $\rm{\tau_{2}^{*}}$ = \SI{3.2}{ms}. In the measurement, the first $\frac{\pi}{2}$ pulse generates a superposition state between two internal ground states of the EF-guided atoms; during the free-evolution time, the two internal ground states enable differential phases, then the second $\frac{\pi}{2}$ pulse causes atomic interference between the two internal ground states. We also conducted frequency-scan Ramsey interferometry [\cref{fig:coherence_793_937_Microwave}b(ii)], sweeping the resonant detuning of the microwave pulses ($\omega_0 + \delta\omega$) for a fixed T to measure the atom interferometric fringe as $\frac{\pi}{2} (\omega_0 + \delta\omega)$ $\rightarrow$ $\rm{T}$ $\rightarrow$ $\frac{\pi}{2}(\omega_0 + \delta\omega)$. 
The amplitude envelope is indicated by the $F (\delta\omega, \Omega, T)$ function, which includes a sinusoidal atom-interferometric fringe under the upper $\rm{sinc^2}$ envelope. The Ramsey fringe spacing $\delta\omega_{\rm Ramsey}/2\pi$ is inversely proportional to T, and the width of the Ramsey fringe's amplitude envelope is proportional to the Rabi frequency $\Omega$.

For all coherence experiments, a multi-pulse detection scheme (\cref{fig:setup}b) is employed. The atom detection is performed using an EF-coupled probe with a power of \SI{10}{pW}, which is resonant with the $\ket{F = 4} \rightarrow \ket{F' = 5}$ transition of \Cs{133} atoms at a wavelength of 852 nm. The transmitted beam power is detected using an SPCM and the counts are recorded in five steps: (1) to detect the atoms in the state $\ket{F = 4}$, the probe pulse is switched on for \SI{1}{ms}; (2) a \SI{100}{\micro s} repump pulse is used to transfer all atoms in the lower ground state $\ket{F = 3}$ into $\ket{F = 4}$; (3) to detect all atoms in the EF atom guide, the probe pulse is switched back on for \SI{1}{ms}; (4) to release all atoms from the EF atom guide, the red-detuned evanescent traveling wave is switched off for a long (\SI{10}{ms}) interval; and (5) a \SI{1}{ms} probe pulse is measured and used as a reference.

This method allows for normalized detection that divides out the noise during detection. To analyze the data, we assume that the number of atoms in the EF atom guide is constant and enumerate the counts detected for the three probe pulses as $c_1, c_2$, and $c_3$, respectively. The transmission during the first probe pulse is $T_1 = c_1/c_3$ and the transmission during the second probe pulse is $T_2 = c_2/c_3$. The number of atoms in the $\ket{F = 4}$ state is proportional to the absorption $A_1 = 1 - T_1$ during the first probe pulse, and the total number of atoms is proportional to the absorption during the second probe pulse $A_2 = 1 - T_2$. Finally, the probability of the atoms being in $\ket{F = 4}$ can be expressed as $P_4 = \frac{c_3 - c_1}{c_3 - c_2}$.

\subsection*{Atomic Coherence Measurements Using EF-Coupled Doppler-Free Raman Beams in 793/937-nm EF Atom Guides on an Optical Nanofiber}

\begin{figure}[t!]
\includegraphics[width=1\columnwidth]{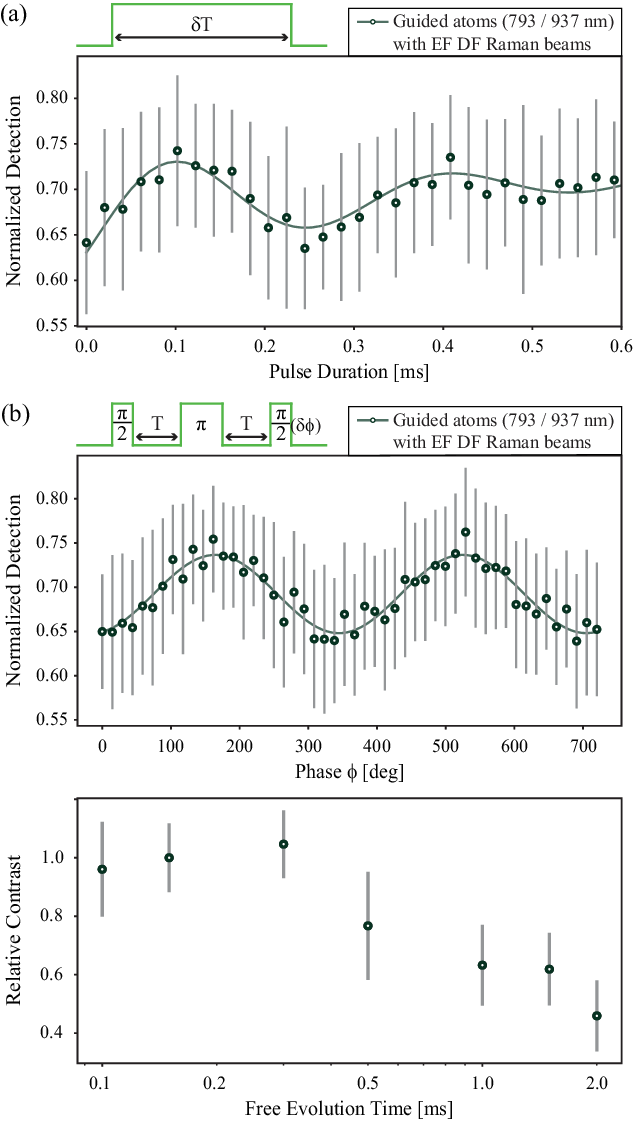}
\caption{Measurements of atomic coherence in EF-guided atoms using EF co-propagating Doppler-free Raman beams. The optical nanofiber testbed for EF atom guides employs light wavelengths of 793 nm and 937 nm. (a) Rabi oscillation measurement; the $1/e$ decay time is $\tau_{1/e}$ = \SI{0.3}{ms}. (b) Ramsey coherence measurements. (i) Phase-scan Ramsey measurement, where the Ramsey sequence with an echo is $\frac{\pi}{2} \rightarrow$ T $\rightarrow$ $\pi$ $\rightarrow$ T $\rightarrow \frac{\pi}{2} (\delta \phi)$, the free evolution time in Ramsey is T = \SI{150}{\micro s}, and the $\pi$ pulse time is \SI{128}{\micro s}. The normalized detection is based on $\ket{F=3}$ state. (ii) Relative Ramsey fringe contrast versus T, varied from \SI{100}{\micro s} to \SI{2}{ms}, normalized to the data in panel (b)(i) at T = \SI{150}{\micro s}. In all panels, each data point is the average of 30 measurements.}
\label{fig:coherence_793_937_DF_Raman}
\end{figure}

Next, we validate the atomic coherence of EF-guided atoms using EF-coupled Raman beams to drive Doppler-free Raman transitions between the atomic clock states (\cref{fig:coherence_793_937_DF_Raman}). The EF beam propagates as the fundamental $\textrm{HE}_{11}$ mode and, similar to the trapping beams, exhibits both radial and azimuthal dependencies in EF intensity. A comparable configuration was implemented in \cite{Polzik17}; however, to our knowledge, our atomic coherence measurement with EF-coupled Doppler-free Raman beams is the first demonstration for EF atom guides.

To generate the EF-coupled Doppler-free Raman beams, we coupled a single phase-modulated beam to the optical nanofiber. An electro-optic modulator (EOM), modulated at the hyperfine frequency of \SI{\sim 9.2}{GHz}, was employed to phase-modulate the beam. The carrier and -1 sideband were utilized to drive the transitions. The carrier was offset locked to the repump transition at \SI{-3.2}{GHz} using a Vescent D2-135 offset lock servo. The EF-coupled Doppler-free Raman beams encompass both transverse and longitudinal polarization components. The quasi-linear polarization of the EF-coupled Raman beams (\cref{fig:setup}c, right) was aligned parallel to both the blue- and red-detuned trap fields (\cref{fig:setup}c, left). When we established a quantization axis along the x-direction (\cref{fig:setup}c, right), the Raman field polarizations above and below the optical nanofiber were $\sigma^{+}$ and $\sigma^{-}$, respectively.\cite{Mitsch14, Meng18, Polzik17} 

\cref{fig:coherence_793_937_DF_Raman}a shows the Rabi oscillations observed when using \SI{\sim 150}{nW} of optical power while scanning the pulse duration. The Raman pulses occur within the physics measurement region (\cref{fig:setup}b, green region). The quantization axis is aligned along the x-axis (\cref{fig:setup}c, right). For reference, the quantization axis of optical pumping is along the y-axis (\cref{fig:setup}c, middle). To facilitate the Raman transitions, we adiabatically transfer the quantization axis from the y-axis to the x-axis within \SI{400}{\micro s}.

For Ramsey coherence measurements (\cref{fig:coherence_793_937_DF_Raman}b), we applied Doppler-free Raman beams to EF-guided atoms in the light-pulse sequence of $\frac{\pi}{2} \rightarrow$ T $\rightarrow$ $\pi$ $\rightarrow$ T $\rightarrow \frac{\pi}{2}$, where the $\pi$ pulse duration is \SI{128}{\micro s}. By varying the phase of the last pulse [\cref{fig:coherence_793_937_DF_Raman}b(i)], we observed a sinusoidal signal, which we fit using the equation $A \sin(\phi+\phi_{\rm offset})+ B$, where A is the fringe contrast and B the offset. To assess the effect of longer free evolution time T between the pulses, we varied $\textrm{T}$ and recorded the contrast [\cref{fig:coherence_793_937_DF_Raman}b(ii)]. We observe that the contrast decays to about \SI{50}{\%} of its value at T = \SI{150}{\micro s} [\cref{fig:coherence_793_937_DF_Raman}b(i)] when T = \SI{2}{ms}.

\subsection*{Atomic Coherence Measurements using Microwave Fields in 685/937-nm EF Atom Guides on an Optical Nanofiber }

To benchmark our heat-efficient 793/937-nm magic-wavelength EF atom guides against a conventional alternative, we performed microwave coherence measurements on a 685/937-nm EF atom guide on an optical nanofiber testbed using traveling evanescent waves in a lin$\parallel$lin polarization configuration. These wavelengths correspond to magic wavelengths for the 6S$_{1/2}$-to-6P$_{3/2}$ D2 transition of \Cs{133}, minimizing light shift effects and facilitating the loading of laser-cooled atoms into the EF guides. With a total power of \SI{23.3}{mW} ($P_{685}$ = \SI{21.8}{mW}, $P_{937}$ = \SI{1.5}{mW}; see Case 3 of \cref{tab:trap_comp}), we obtained N$_{\rm Trap}$ = 60\SI{\pm 4}{} EF-guided atoms on an optical nanofiber testbed. The lifetime of the EF-guided atoms was measured at $\tau_{\rm Trap}$ = 14.0\SI{\pm 0.6}{ms}, compared to 5.2\SI{\pm 0.8}{ms} for the background atoms. The nanofiber diameter was $\rm{D_{NF}}$ = \SI{420}{nm}, with optical potential minima located approximately \SI{260}{nm} from the nanofiber surface, aligned with the polarization direction.

\begin{figure}[t!]
\includegraphics[width=1\columnwidth]{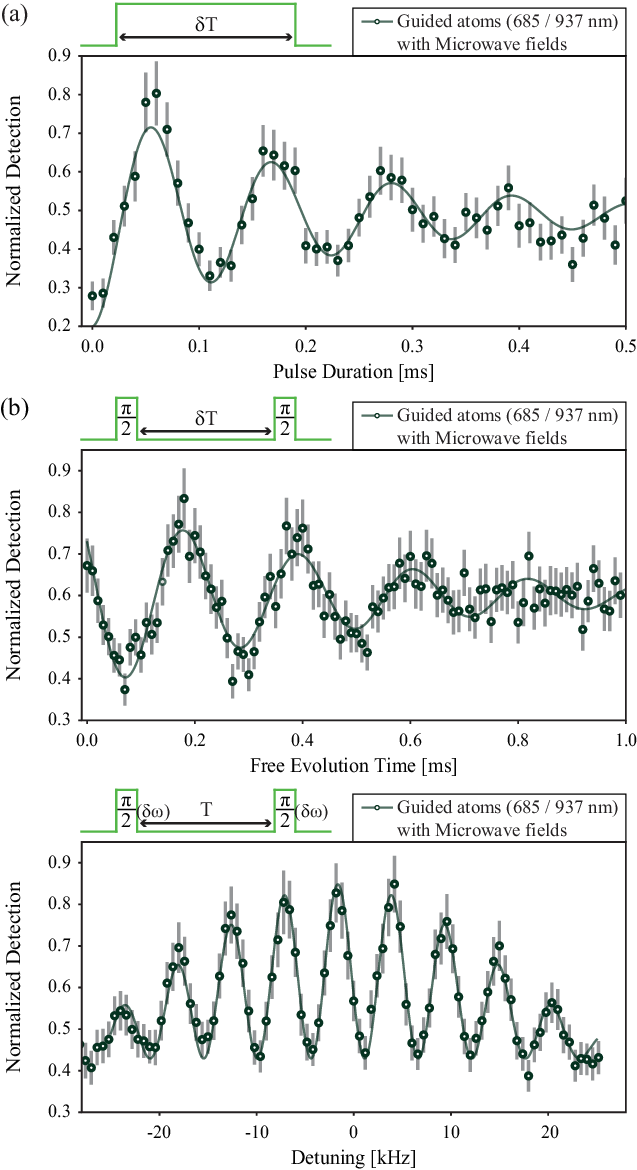}
\caption{Measurements of atomic coherence in EF-guided atoms using microwave fields. The optical nanofiber testbed for EF atom guides employs light wavelengths of 685 nm and 937 nm. (a) Rabi oscillation measurement; the $1/e$ decay time is $\tau_{1/e}$ = 220\SI{\pm 3}{\micro s}. (b) Ramsey coherence measurements. (i) Time-scan Ramsey measurement, where the Ramsey sequence is $\frac{\pi}{2} \rightarrow$ $\delta$T $\rightarrow \frac{\pi}{2}$, the coherence time of EF-guided atoms is $\tau_2^*$ = 470\SI{\pm 60}{\micro s}, and the $\pi$ pulse time of the microwave is \SI{55}{\micro s}.  (ii) Frequency-scan Ramsey measurement, where the Ramsey sequence is $\frac{\pi}{2}$ ($\delta\omega$) $\rightarrow$ T $\rightarrow$ $\frac{\pi}{2}$ ($\delta\omega$) and the free-evolution time is T = \SI{150}{\micro s}. In all panels, each data point is the average of 100 measurements.}
\label{fig:coherence_685_937_Microwave}
\end{figure}


We investigated the atomic coherence of EF-guided atoms by driving transitions between the atomic clock states, $\ket{F = 3, m_F = 0}$ to $\ket{F = 4, m_F = 0}$, using a microwave horn. The microwave field requires a well-defined quantization axis (\cref{fig:setup}c, middle). In the Rabi oscillation measurement (\cref{fig:coherence_685_937_Microwave}a), the microwave field is nearly resonant with the atomic clock transition (\SI{\sim 9.192}{GHz}), revealing coherent Rabi oscillations as the pulse length increases.

For the time-scan Ramsey measurement [\cref{fig:coherence_685_937_Microwave}b(i)], we extended the free-evolution time ($\rm{T}_0 + \delta T$) to measure atom-interferometric fringes using the pulse sequence $\frac{\pi}{2}$ $\rightarrow$ $\rm{T}_0 + \delta T$ $\rightarrow$ $\frac{\pi}{2}$. The first $\frac{\pi}{2}$ pulse creates a superposition state, and the second pulse induces interference. We also performed frequency-scan Ramsey interferometry [\cref{fig:coherence_685_937_Microwave}b(ii)], sweeping the microwave detuning ($\omega_0 + \delta\omega$) to measure the fringe. These experiments utilized a three-pulse detection scheme (\cref{fig:setup}b), where atom detection was conducted with an EF-coupled probe at \SI{10}{pW}, resonant with the $\ket{F = 4} \rightarrow \ket{F' = 5}$ transition of \Cs{133} atoms at 852 nm. 

Compared to the 793/937-nm EF atom guides, the 685/937-nm EF atom guides require higher optical power (see \cref{tab:trap_comp}) and incur larger bending losses in membrane-waveguide PIC platforms due to the greater wavelength separation. Our choice of magic-wavelength pairs considers the specific waveguide geometries needed for compact, low-SWaP quantum accelerometers and gyroscopes.

\section*{Discussion and Outlook}

To minimize lateral atomic motion and ensure transverse confinement for EF-guided atom interferometry, we designed and fabricated 793/937-nm EF atom guides (with both linear and racetrack geometries, \cref{fig:protocol} \& \cref{fig:fab_device}) on alumina membrane-waveguide PIC platforms. Because direct atom trapping using only EF-coupled beams on PIC waveguides remains challenging, we evaluated our proposed 793/937-nm light configuration (magic wavelengths for the \Cs{133} D2 line at 852 nm) using optical nanofiber testbeds. On these testbeds, we demonstrated a low-power (\SI{\sim 5}{mW}) EF atom guide, characterized the atom number and lifetime of EF-guided atoms, and confirmed their coherence via microwave fields and EF-coupled, \SI{150}{nW} Doppler-free Raman beams. These results represents a critical first step toward full EF-guided atom interferometry and on-chip quantum inertial sensors.

Ongoing efforts will exploit state-dependent photon recoils of EF-guided atoms in the nanofiber testbeds to realize proof-of-concept acceleration sensing with EF-coupled, Doppler-sensitive Raman beams. We also anticipate that advanced cooling schemes and large-momentum-transfer techniques will further boost the sensitivity of EF-guided atom interferometers. Meanwhile, our recent progress in membrane-waveguide PIC platforms \cite{Gehl21, Lee21}  offers improved scalability and design flexibility, reduced SWaP, efficient generation of dense cold atom ensembles for loading into EF atom guides, and superior thermal management. Altogether, the benchmarked performance on optical nanofibers paves the way for fully integrated, multi-axis, low-SWaP quantum inertial sensors on membrane-waveguide PIC platforms.\cite{Lee24a, Lee24b, Lee25} 

For future atom interferometry demonstrations on the linear nanofiber testbed, the light polarization must be chosen to support coherent matterwave splitting and recombination. In particular, Doppler-sensitive Raman transitions require counter-propagating fields to impart state-dependent momentum kicks. In an optical nanofiber testbed, these fields acquire a propagation-dependent polarization because of the longitudinal component of the fundamental $\textrm{HE}_{11}$ mode.\cite{Mitsch14}  For example, two EF-coupled Raman fields that are quasi-linearly polarized along x and counter-propagating along +z and -z, with the quantization axis along y, will become circularly polarized with opposite handedness at the +x versus -x positions (\cref{fig:setup}c, right).\cite{Mitsch14}  We exploit this effect to drive atom interferometry using magnetically sensitive internal states.\cite{Bernard22} When atoms are prepared in the state $\ket{F, m_F=-1}$, two counter-propagating Raman beams (one  $\sigma^{+}$ and the other $\sigma^{-}$) induce a two-photon transition with $\Delta m_F = 2$  to $\ket{F, m_F=+1}$, which is first-order insensitive to magnetic fields and has been used to realize free-space light-pulse atom interferometers.\cite{Bernard22} Recent work by \cite{Pennetta25} demonstrated nanofiber-based hybrid trapping via a traveling-wave, blue-detuned beam under the van der Waals potential. The resulting extended atomic coherence times and trap lifetimes are expected to benefit the ongoing development of on-chip quantum inertial sensors using membrane-waveguide PIC platforms.

\section*{Materials and Methods}

\subsection*{Fabrication of Alumina Membrane-Waveguide PIC Platforms}

\begin{figure}[t!]
\includegraphics[width=0.9\columnwidth]{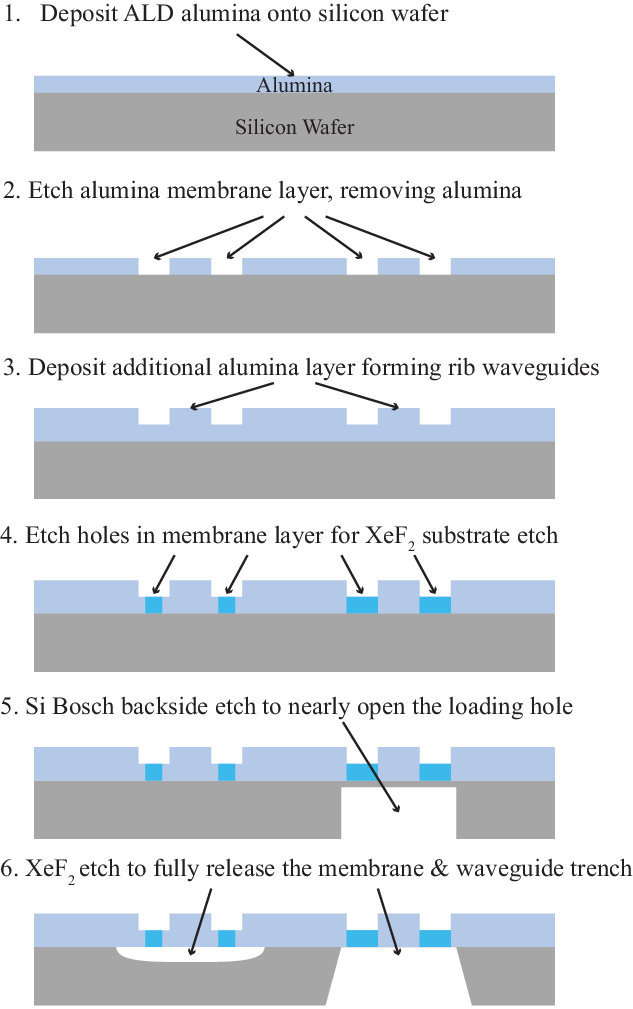}
\caption{Fabrication process of the membrane-waveguide PIC platform for EF atom guiding: (1) deposit a thin alumina film on a Si wafer by atomic layer deposition (ALD); (2) selective etch to create thin-membrane or membrane-free regions.; (3) deposit a second alumina film by ALD to form waveguides; (4) pattern holes in this layer for localized substrate removal during the $\rm{XeF}_{2}$ etch; (5) etch a window through the backside of the Si wafer for optical access; and (6) perform a final $\rm{XeF}_{2}$ etch to fully suspend the waveguides and open the access hole.}
\label{fig:fab_process}
\end{figure}

Membrane-waveguide PIC platforms (\cref{fig:fab_device}), utilizing atomic layer deposition (ALD) to deposit thin-films of alumina ($\rm{Al_{2}O_{3}}$; $\rm{n_{\rm{Al_{2}O_{3}}}=1.76}$), exhibit excellent waveguide properties from UV to NIR wavelengths, with enhanced resistance to alkali vapor. This work presents devices with suspended membrane waveguides, enabling spans greater than \SI{> 1}{cm} (\cref{fig:fab_device}e) without substrate loss by locally removing the substrate around the waveguide. Span length is limited by heat generation in the waveguide due to optical losses, which have been previously measured at approximately \SI{1}{dB/cm} loss in similar structures. 

The fabrication process consists of six steps (\cref{fig:fab_process}): (1) performing ALD of a thin alumina film on a silicon wafer; (2) patterning the alumina film into ridge-templates for waveguides by photolithography and inductively coupled plasma reactive-ion etching (ICP-RIE); (3) performing ALD of a second alumina film to coat those ridges and form the waveguide structure; (4) etching openings for substrate removal and creating large optical access windows; (5) opening the backside windows using deep reactive-ion etching (DRIE), stopping \SI{< 50}{\micro m} from the alumina; and (6) performing selective silicon $\rm{XeF}_{2}$ etching to fully open the windows and suspend the waveguide through adjacent membrane openings, resulting in a \SI{\sim 50}{\micro m} trench.

\subsection*{Fabrication Process and Characterization of Optical Nanofibers}

\begin{figure}[t!]
\includegraphics[width=0.9\columnwidth]{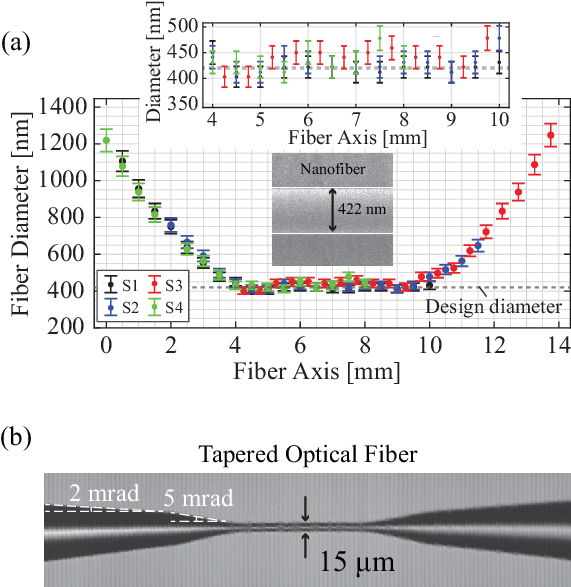}
\caption{Manufactured optical nanofibers for linear EF atom guides. (a) Scanning electron microscope (SEM) measurements confirm nanofiber diameters, which were previously designed at \SI{420}{nm} using algorithmic fiber-pulling. (Inset) Close-up SEM image of an optical nanofiber. (b) Optical microscope images of a \SI{15}{\micro m}-diameter tapered optical fiber with linear and exponential sections.}
\label{fig:NF_fab}
\end{figure}

The optical nanofiber (see \cref{fig:NF_fab}a-c) was fabricated from single-mode fiber (780 HP) using a stationary oxy-hydrogen torch and two motor stages to create linear and exponential-tapered fiber sections. An algorithmic fiber-pulling method \cite{Hoffman14, Ward14} effectively reduced the fiber diameter from 125 to less than \SI{0.5}{\micro m}, optimizing taper lengths and waist diameters.

To achieve over \SI{99.5}{\%} transmission between the unmodified fiber mode (LP$_{01}$) and the EF mode (HE$_{11}$), the fiber has two symmetric linear tapers (\SI{2}{mrad} angle, 2$\times$\SI{2.882}{cm}, from \SI{125}{\micro m} to \SI{12}{\micro m} diameter) and two symmetric exponential tapers (2$\times$\SI{1.113}{cm}, from 12 to less than \SI{0.5}{\micro m} diameter), totaling \SI{8.477}{cm}. To enhance atom-light interaction efficiency, the nanofiber waist diameter ($\rm{D_{NF}}$) was set to \SI{420}{nm}; to optimize rigidity and functionality in EF-guided atom interferometry, the section length was set to \SI{5}{mm} .

When the fiber-pulling process was complete, the tapered optical fiber was mounted onto a 3D-printed titanium mount using ultra-violet (UV) epoxy. The mount was then placed into a vacuum chamber (see \cref{fig:setup}a) via an extended hollow adapter connected through a groove grabber to a stainless-steel chamber. The diameter of the nanofiber section (420\SI{\pm 10}{nm}) was measured using a scanning electron microscope (SEM) across four samples, confirming the consistency of this fabrication method (see \cref{fig:NF_fab}b). Preliminary tests indicated that the optical nanofibers can endure over \SI{150}{mW} of optical power without damage at a vacuum level of $10^{-8}$ mbar, which is adequate for demonstrating EF atom guides.

Background atoms reduce probe transmission from \SI{99}{\%} to \SI{36}{\%} over \SI{4}{s} due to atom adsorption onto the nanofiber surface. The shot-to-shot measurement cycle (\SI{< 0.5}{s}) allows us to clean the fiber by turning on the 937 nm laser at the end of each run.

\begin{acknowledgements}
We would like to express our gratitude to Craig W. Hogle, Jonathan Sterk, and Weng Chow for their support and helpful discussions. This work was supported by the Laboratory Directed Research and Development program at Sandia National Laboratories and has funding under the DARPA APhI program. Sandia National Laboratories is a multimission laboratory managed and operated by National Technology and Engineering Solutions of Sandia, LLC., a wholly owned subsidiary of Honeywell International, Inc., for the U.S. Department of Energy's National Nuclear Security Administration under contract DE-NA-0003525. 
\end{acknowledgements}

\section*{Author Declarations}
\subsection*{Conflict of interest}
The authors have no conflicts to disclose.

\section*{Data Availability}
The data that support the findings of this study are available from the corresponding author upon reasonable request.


\end{document}